# Cohesive energy of zinc blende ($A^{III}B^{V}$ and $A^{II}B^{VI}$) structured solids

**A. S. Verma**[*,1], B. K. Sarkar[2] and V. K. Jindal[1]

[1]Department of Physics, Panjab University, Chandigarh, (India), 160014
[2]Department of Physics, VIT University, Vallore, Tamil Nadu (India), 632014

**Abstract:** In this letter we present an expression relating the cohesive energy ($E_{coh}$ in kcal/mol) for the $A^{III}B^{V}$ and $A^{II}B^{VI}$ semiconductors with the product of ionic charges ($Z_1Z_2$) and nearest neighbor distance d (Å). The cohesive energy values of these solids exhibits a linear relationship when plotted on a log-log scale against the nearest neighbor distance d (Å), but fall on different straight lines according to the ionic charge product of the solids. A good agreement has been found between the experimental and calculated values of the cohesive energy for $A^{III}B^{V}$ and $A^{II}B^{VI}$ semiconductors.

**PACS** 62.20.-x

[**keywords:** cohesive energy, ionic charge, transition metal chalcogenide and pnictides]

## 1 Introduction

Cohesive energy is one of the parameter in understanding the nature of chemical bonding and several important parameters can be predicted by using it. Its magnitude tells us about the stability and chemical reactivity of solids. Eventually, it is the quantity which determines the structure, because different possible structures would have different cohesive energies [1]. Semi-empirical molecular orbitals have been widely and successfully used to develop the theory of the solid state. Recently [2-6]' frequent attempts have been made to understand the electronic, mechanical, elastic and optical properties of zinc blende ($A^{II}B^{VI}$ and $A^{III}B^{V}$) semiconductors. This is because of their interesting semi-conducting

•Corresponding author: e-mail: ajay_phy@rediffmail.com, mobile: +91 9412884655, Ph. +91 565 2423417

properties and various practical applications in the field of non-linear optics, electronics, photovoltaic detectors, light emitting diodes and solar cells etc. In modern high-speed computer techniques, they allow researchers to investigate many structural and physical properties of materials only by computation or simulation instead of by traditional experiments. Empirical relations have become widely recognized as the method of choice for computational solid-state studies. In many cases empirical relations do not give highly accurate results for each specific material, but they still can be very useful. Empirical concepts such as valence, empirical radii, electronegativity, ionicity and plasmon energy are useful for the evaluation of solid state properties of solids [1,7]. These concepts are directly associated with the character of the chemical bond and thus provide means for explaining and classifying many basic properties of molecules and solids. Any change in crystallographic environment of an atom is related to core electrons via the valence electrons. The change in wave function that occurs for the outer electrons usually means a displacement of electric charge in the valence shell so that the interaction between valence, shell, and core electrons is changed. This leads to a change in binding energy of the inner electron and to a shift in the position of the absorption edge.

In the previous research [8-12], we have calculated the electronic, mechanical and optical properties of binary and ternary semiconductors with the help of ionic charge theory. This is due to the fact that the ionic charge depends on the number of valence electrons, which changes when a metal forms a compound. Therefore we thought it would be of interest to give an alternative explanation for cohesive energy ($E_{coh}$ in kcal/mol) of zinc blende ($A^{III}B^{V}$ and $A^{II}B^{VI}$) structured solids.

## 2 Theory, results and discussion

Aresti et al [13] have studied the cohesive energy of zinc blende solids and proposed an empirical relation for cohesive energy in terms of nearest neighbour distance (d) as follows,

$$E_{coh} = E_{coh}\,(IV) - B(d, R)\,\{1 - \Sigma\, E_{coh}\,(i) / E_{coh}\,(IV)\} \tag{1}$$

Where $E_{coh}$ (IV) is cohesive energy of purely covalent crystals and $B(d, R) = E_{coh}$ (IV) – $k(R)\times d(BX)/d$ is now a parameter depending on d and R.

$$k(R) = C \exp(-Z^{1/2} / 4) \tag{2}$$

where C is constant, which depends the rows and Z = Z(A) + Z(B), atomic number of atom A and atom B.

Recently [14-16] much type of theoretical approaches have been reported to determine the value of cohesive energy of solid-state compounds. H. Schlosser [17,18], has studied the cohesive energy trends in rocksalt structure in terms of nearest neighbour distance using the following relation,

$$E_{coh} = \text{constant} / d \tag{3}$$

In a previous work [11], we proposed a simple relation for dielectric constant of chalcopyrite structured solids in terms of the product of ionic charges and nearest neighbour distance by the following relation,

$$\text{Dielectric constant } (\varepsilon_\infty) = K\,(Z_1 Z_2)^S\, d^2 \tag{4}$$

Where $Z_1$ and $Z_2$ are the ionic charge on the cation and anion respectively, d is the nearest neighbour distance in Å and K and S are constants. The cohesive energy of $A^{III}B^{V}$ and $A^{II}B^{VI}$ semiconductors exhibit a linear relationship when plotted against nearest-neighbour distance, but fall on different straight lines according to the ionic charge product of the compounds, which is presented in figure 1. We observe that in the plot of cohesive energy and nearest neighbour distance; the $A^{III}B^{V}$ semiconductors lie on line nearly parallel to the line for the $A^{II}B^{VI}$ semiconductors. The Krishnan–Roy theory [19], Jayaraman *et al* [20] and Sirdeshmukh *et al* [21] found that substantially reduced ionic charges must be used to get better agreement with experimental values. To obtain better agreement between experimental and theoretical data for zinc blende type crystal structure compounds, Schlosser's relation (3) may be extended to,

$$E_{coh} = \text{constant}\,(Z_1Z_2)^{0.4}/d^{2.5} \tag{5}$$

$Z_1$ and $Z_2$ are ionic charge of cation and anion respectively and d is the nearest neighbour distance in Å. The value of constant for zinc blende type crystal structure is 710. The value of product of ionic charge is 4 for $A^{II}B^{VI}$ and 9 for $A^{III}B^{V}$ semiconductors [9]. A detailed discussion of cohesive energy for these solids has been given elsewhere [13-18] and will not be presented here. The proposed empirical relation (5) has been applied to evaluate for $A^{II}B^{VI}$ and $A^{III}B^{V}$ semiconductors. The results are presented in table 1. The calculated values are in good agreement with the experimental and theoretical values reported by earlier researchers [13].

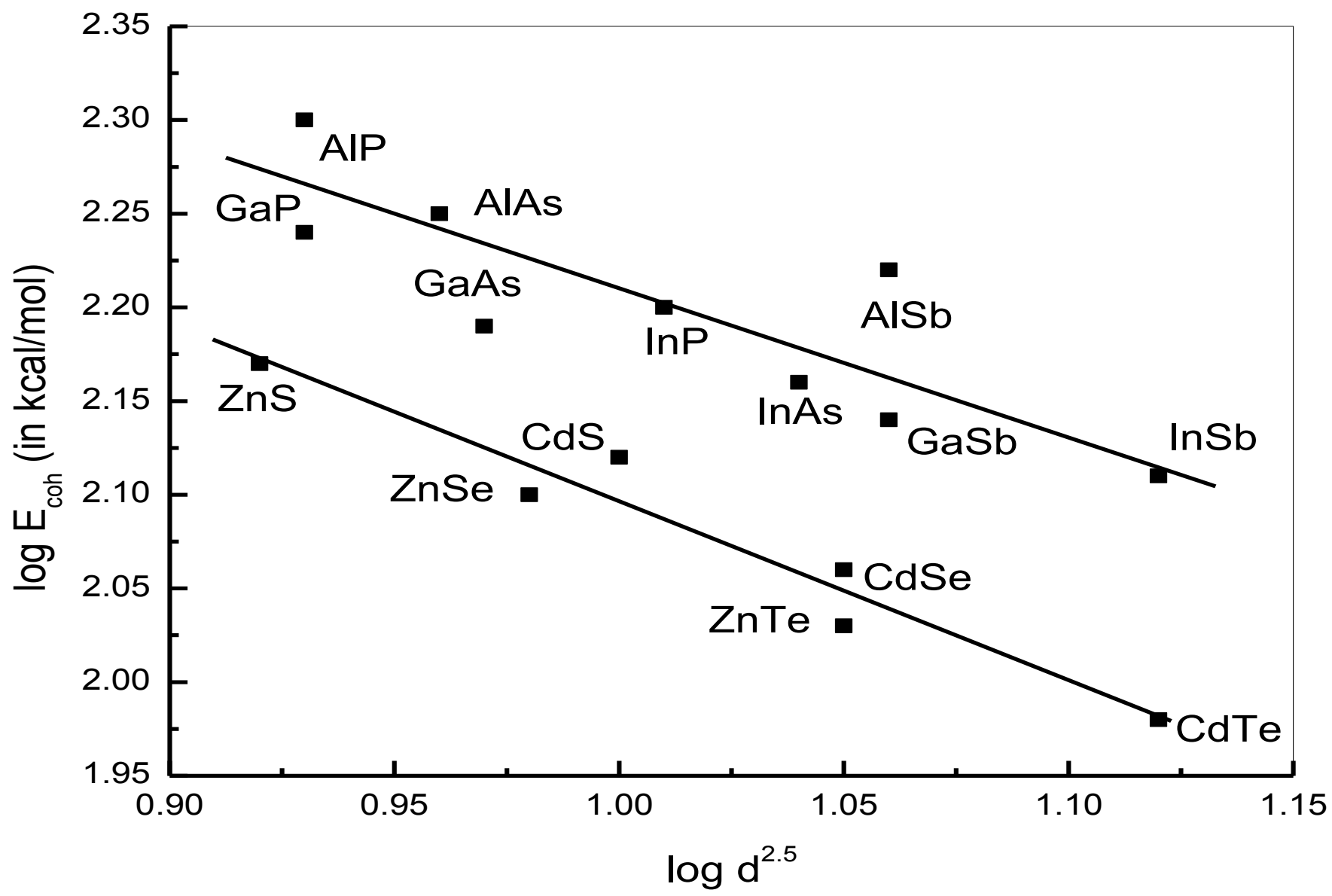

**Figure 1**. Plot of log E (kcal/mol) against log $d^{2.5}$ (nearest neighbour distance in Å) for transitional metal chalcogenides and pnictides. Plots of $A^{II}B^{VI}$ and $A^{III}B^{V}$ semiconductors are nearly parallel. All experimental values are from [13].

**Table 1** Values of cohesive energy ($E_{coh}$ in kcal/mol) for compound ($A^{III}B^{V}$ and $A^{II}B^{VI}$) semiconductors.

| Solids | d | $E_{coh}$ exp. [13] | $E_{coh}$ theor. [13] | $E_{coh}$ [This work] | %error (with experimental) |
|---|---|---|---|---|---|
| ZnS | 2.34 | 146.6 | 151.3 | 147.6 | 0.7 |
| ZnSe | 2.46 | 124.5 | 123.7 | 130.2 | 4.6 |
| ZnTe | 2.64 | 106.3 | 108.6 | 109.2 | 2.7 |
| CdS | 2.52 | 131.6 | 134.6 | 122.6 | 6.8 |
| CdSe | 2.62 | 113.6 | 110.3 | 111.3 | 2 |
| CdTe | 2.81 | 95.8 | 95.4 | 93.4 | 2.5 |
| HgS | 2.53 | | | 121.4 | - |
| HgSe | 2.63 | | | 110.2 | - |
| HgTe | 2.80 | | | 94.2 | - |
| AlP | 2.36 | 198.0 | 197.0 | 199.8 | 0.9 |
| AlAs | 2.43 | 178.9 | 177.2 | 185.8 | 3.9 |
| AlSb | 2.66 | 165.0 | 162.4 | 148.2 | 10 |
| GaP | 2.36 | 173.8 | 173.2 | 199.8 | 15 |
| GaAs | 2.45 | 154.7 | 154.6 | 181.9 | 17 |
| GaSb | 2.65 | 138.6 | 140.5 | 149.6 | 7.9 |
| InP | 2.54 | 158.6 | 159.3 | 166.3 | 4.9 |
| InAs | 2.61 | 144.3 | 141.7 | 155.4 | 7.7 |
| InSb | 2.81 | 128.5 | 128.3 | 129.2 | 0.5 |
| BAs | 2.04 | | | 287.7 | - |
| BSb | 2.24 | | | 227.7 | - |
| TiP | 2.49 | | | 175.8 | - |
| TiAs | 2.58 | | | 159.9 | - |
| TiSb | 2.75 | | | 136.3 | - |

## 3 Conclusion

In conclusion, we have presented an empirical expression relating cohesive energy with the product of ionic charges and nearest neighbour distance. The values obtained from the expression for II-VI and III-V semiconductors agree well with reported experimental values.

**Acknowledgements**: One of the authors (Dr. Ajay Singh Verma, PH/08/0049) is thankful to the University Grant Commission New Delhi, India for supporting this research under the scheme of U.G.C. Dr. D. S. Kothari Post Doctoral Fellowship.

## References

[1] L. Pauling, The Nature of the Chemical Bond, 3rd, ed. (Cornell University Press, Ithaca, 1960).
[2] Y. Al-Douri, H. Abid and H. Aourag, Materials letters, **59**, 2032, (2005).
[3] A. E. Merad, M. B. Kanoun, G. Merad, J. Cibert and H. Aourag, Materials Chem. and Phys, **92**, 333, (2005).
[4] H. M. Tutuncu, S. Bagci, G. P. Srivastava, A. T. Albudak and G. Ugur, Phys. Rev. B, **71**, 195309, (2005).
[5] S. Q. Wang and H. Q. Ye, J. Phys: condensed matter, **17**, 4475, (2005).
[6] S. H. Sohn, D. G. Hyun, M. Noma, S. Hosomi and Y. Hamakawa, J. Crytal Growth, **117**, 907, (1992).
[7] V. Kumar, and B. S. R. Sastry, J. Phys. Chem. Solids, **66**, 99 (2005).
[8] A. S. Verma, Physica Scripta, **79,** 045703, (2009).
[9] A. S. Verma, Physics letters A, **372**, 7196, (2008).
[10] A. S. Verma, Physica Status Solidi B, **246**, 192, (2009).
[11] A. S. Verma, Solid State Communications, **149**, 1236, (2009).
[12] A. S. Verma, Philosophical Magazine, **89**, 183, (2009).
[13] A. Aresti, L. Garbato and A. Rucci, J. Phys. Chem. Solids, **45**, 361, (1984).
[14] L. M. Liu, S. Q. Wang and H. Q. Ye, J. Phys. condensed matter, **17**, 5335 (2005).
[15] J. Haglund, G. Grimvall, T. Jarlborg and A. F. Guillermet, Phys. Rev. B, **43,** 14400, (1991).
[16] R. C. Mota, S. C. Costa, P. S. Pizani and J. P. Rino, Phys. Rev. B, **71,** 224114, (2005).
[17] H. Schlosser, J. Phys. Chem. Solids, **53**, 855, (1992).
[18] H. Schlosser, Phys. Status Solidi b, **179**, k1, (1993).
[19] K. S. Krishnan and S. K. Roy, Proc. R. Soc. London, **210**, 481, (1952).
[20] A. Jayaraman, B. Batlogg, R. G. Maines and H. Bach, Phys. Rev. B, **26**, 3347, (1982).
[21] D. B. Sirdeshmukh and K. G. Subhadra, J. Appl. Phys., **59**, 276, (1986).